\documentclass[12pt]{article}
\begin{document}
\begin{titlepage}
\null
\vskip 1cm
\begin{center}{\Large Dynamical effects of a cosmological constant}\\
\vskip 1cm {\bf Lars Bergstr\"om}\\
\vskip .2cm
{\em Department of Physics,
Stockholm University,\\
         Box 6730, SE-113 85 Stockholm, Sweden,}\\ {\tt lbe@physto.se}\\

\vskip .2cm
and\\
\vskip.2cm
{\bf Ulf Danielsson}\\
\vskip .2cm
{\em Department of Theoretical Physics,
Uppsala University,\\
        Box 807, SE-751 08 Uppsala, Sweden,}\\ {\tt ulf@teorfys.uu.se}
\end{center}
\vskip .2cm
\centerline{{\bf \today}}

\begin{abstract}
The observational evidence for the existence of a non-zero cosmological
constant is getting stronger. It is therefore timely to address the
question of its eventual effect on the dynamics of galaxies, clusters
and larger structures in the Universe. We find, contrary to a recent
claim, that the influence of the cosmological constant has to be
negligible for, e.g., the rotation curves of galaxies. On larger scales,
starting with large galaxy clusters, there are potentially measurable
effects from the repulsive addition to the Newtonian gravitational force
caused by the cosmological constant.
\end{abstract}
\end{titlepage}

During the past few years, remarkable progress has been made in cosmology,
both observational and theoretical. One of the outcomes of these rapid
developments is the increased confidence that most of the energy density of
the observable universe is of an unusual form, i.e., not made up of the
ordinary matter (baryons and electrons) that we see around us in our
everyday world.

There are convincing arguments for the existence of a large amount of
non-luminous, i.e., dark, matter. The matter content of the universe is at
least a factor of 5 higher than the maximum amount of baryonic matter
implied by big bang nucleosynthesis. This dark matter is thus highly likely
to be ``exotic'', i.e, non-baryonic.

There are also indications, although still not entirely conclusive, of the
existence of vacuum energy, corresponding to the famous ``cosmological
constant'' that Einstein introduced but later rejected (although without
very good reasons) in his theory of general relativity. This possibility has
recently been given increased attention due to results from Type Ia
supernova surveys \cite{ariel,kirshner2}. (For a recent review of the
observational status of dark matter and dark energy, see \cite{triangle}.)

It may be interesting to investigate possible consequences of the
cosmological constant besides its influence on the geometry of the universe,
and the redshift-dependence of the luminosity-distance relation for standard
candles \cite{BGbook}. This is the subject of the present paper. We find
disagreement with a recent paper \cite{kraniotis} where the cosmological
constant was claimed to influence the rotation curves of galaxies strongly.
However, some small effects on the dynamcis of galaxy clusters do not seem
excluded.

Let us first set our conventions. Einstein's equations read 
\begin{equation}
R^{\mu \nu }-{\frac{1}{2}}g^{\mu \nu }-\Lambda g^{\mu \nu }=8\pi G_{N}T^{\mu
\nu }.
\end{equation}
The energy density in the form of a co{smological constant $\Lambda $ can be
conveniently written in units of the density scaled to the critical density, 
\begin{equation}
\Omega _{\Lambda }\equiv {\frac{\rho _{\Lambda }}{\rho _{\mathrm{crit}}}},
\end{equation}
where 
\begin{equation}
\rho _{\Lambda }={\frac{\Lambda }{8\pi G_{N}}}  \label{eq:rhol}
\end{equation}
with $G_{N}=1/m_{\mathrm{Pl}}^{2}$ (the numerical value of the Planck mass
is $m_{\mathrm{Pl}}=1.2\cdot 10^{19}$ GeV) and the present value of the
critical density 
\begin{equation}
\rho _{\mathrm{crit}}={\frac{3H_{0}^{2}}{8\pi G_{N}}}.
\end{equation}
Thus, 
\begin{equation}
\Omega _{\Lambda }={\frac{\Lambda }{3H_{0}^{2}}}.
\end{equation}
Writing $H_{0}=100h$ km\thinspace s$^{-1}$\thinspace Mpc$^{-1}$ (with $h\sim
0.6\pm 0.1$ from observations), the present numerical value of the critical
density is 
\begin{equation}
\rho _{\mathrm{crit}}\simeq 8\cdot 10^{-47}h^{2}\ \mathrm{GeV}^{4}.
\end{equation}
}

This was derived using particle physics units ($c=\hbar =1$). Expressed in $%
cgs$ units, the presently observationally favoured value \cite{triangle} $%
\Omega _{\Lambda }\simeq 0.7\pm 0.2$ then translates to 
\begin{equation}
\Lambda _{\mathrm{obs}}\simeq 10^{-56}\ \mathrm{cm}^{-2}.
\label{eq:lambda_obs}
\end{equation}
In a recent preprint \cite{kraniotis}, an attempt was made to explain the
flat rotation curves of galaxies as being due to the effect of a
cosmological constant instead of the ``traditional'' explanation in terms of
dark matter. However, values some three to four orders of magnitude 
larger than
that in (\ref{eq:lambda_obs}) were needed, something which is clearly in
extreme disagreement with observations. In fact, there seems to be a further
mistake, a sign error, in \cite{kraniotis}. A positive cosmological
constant, as favored by the observations, will tend to accelerate the
expansion of the universe and, if anything, make matter in the outer
regions of galaxies less rather than
more bound.

While the effects of a cosmological constant thus are negligible on the
length scale of galaxies, one might expect observable consequences for
galaxy clusters. As a first attempt to see such an effect, we will consider
the fate of circular orbits in a flat, expanding universe with a
cosmological constant. To do this we start with the equation of motion for a
particle in an expanding universe with an additional gravitational
potential, 
\[
\ddot{\overline{\chi }}+\frac{2\dot{a}}{a}\overline{\chi }=-\frac{\overline{g%
}}{a},
\]
where $\overline{\chi }$ is the comoving coordinate and $a$ is the scale
factor. In terms of physical distances, $\overline{R}=a\overline{\chi }$,
one finds 
\[
\ddot{\bar{R}}-\bar{R}\frac{\ddot{a}}{a}=-\overline{g}.
\]
As an example we consider the de Sitter case with $\Omega _{\Lambda }=1$,
i.e. a universe totally dominated by the cosmological constant. We then use 
\[
a(t)=e^{Ht}
\]
where $H=\sqrt{\frac{\Lambda }{3}}$ is constant, to obtain 
\begin{equation}
\frac{v^{2}}{R}=\frac{G_{N}m}{R^{2}}-\frac{\Lambda }{3}R.
\label{bandesitter}
\end{equation}
for an object in orbit around a central mass $m$. One might note that the
same result may be obtained by starting with the static form of the de
Sitter metric, i.e. 
\[
ds^{2}=\left( 1-\frac{2G_{N}m}{r}-\frac{\Lambda }{3}r^{2}\right) dt^{2}-%
\frac{dr^{2}}{\left(1-\frac{2G_{N}m}{r}-\frac{\Lambda }{3}r^{2}\right)}-r^{2}d\Omega ^{2},
\]
and using a Newtonian analysis. This form of the metric is related to the
cosmological form through a coordinate transformation. Equation (\ref
{bandesitter}) shows that for large enough $R,$ i.e. 
\[
R>\left( \frac{3G_{N}m}{\Lambda }\right) ^{1/3},
\]
there are no longer bound orbits. Does this have observable consequences?
Unfortunately it is easy to see that the effect becomes important only for
orbits that are such that they have periods of the order of the age of the
universe. Furthermore, the effect of the cosmological constant decreases
rapidly for smaller orbits. Hence the concept of a rotation curve loses its
meaning and we had better look elsewhere for a better approach to the
possible effect of a cosmological constant. One possibility is the way
clusters and superclusters of galaxies form.

As a first step, one may consider the infall of matter onto a galaxy cluster
in the regime where linear perturbation theory is valid. This has in fact
been treated by Peebles \cite{jim}, in the case $\Omega_M+\Omega_\Lambda=1$
(i.e. zero curvature) which is the natural case in view of inflation, and
which, incidentally, is now also indicated by the recent balloon
measurements of the cosmic microwave background \cite{boomerang}. Peebles
showed that, unfortunately, the dependence of the infall peculiar velocity $%
v $ for a cluster of proper radius $R$ and overdensity $\delta$ on $%
\Omega_\Lambda$ is quite weak, being well parametrized by 
\begin{equation}
v=0.3H_0R\delta\Omega_M^{0.6}
\end{equation}
for $0.03 < \Omega_M < 0.3$ and $1< \delta < 3$, essentially independent of $%
\Omega_\Lambda$. This formula was generalized to arbitrary 
$\Omega_M+\Omega_\Lambda$ in \cite{lahav}.

In the non-linear, collapsing phase, we may obtain an estimate of the
influence of $\Lambda$ by adapting the simple constant density, spherical
collapse model \cite{kolbturner} to the situation when the cosmological
constant is present. We thus look at the situation when an overdense region
expands to a maximal radius $R_{\mathrm{max}}$, and then contracts to a
viral radius $R_{\mathrm{vir}}$.

To analyse this situation, we use the equation for the energy per unit mass (%
\emph{cf.} \cite{jim}, Eq.~(20) or our equation (\ref{bandesitter})) of a
mass shell of proper radius $R(t)$ containing a fixed mass $m$: 
\begin{equation}
E={\frac{\dot{R}^{2}}{2}}-{\frac{G_{N}m}{R}}-{\frac{\Lambda R^{2}}{6}},
\end{equation}
where the three terms correspond to the kinetic energy, Newtonian
gravitational energy, and vacuum energy, respectively. As in the standard
analysis \cite{kolbturner}, we may employ the virial theorem relating the
average value of the kinetic energy $T$ to the potential energy $V$ 
\begin{equation}
\langle 2T\rangle =\langle \vec{r}\cdot {\frac{\partial V}{\partial \vec{r}}}%
\rangle .
\end{equation}

Taking the average over a sphere of constant density of the energy equation
(at the turn-around radius $R_{\mathrm{max}}$ the kinetic energy is zero) 
\begin{equation}
E=T_{\mathrm{vir}}+V^G_{\mathrm{vir}}+V^\Lambda_{\mathrm{vir}}=V^G_{\mathrm{%
max}}+V^\Lambda_{\mathrm{max}},
\end{equation}
utilizing 
\begin{equation}
{\langle \Lambda r^2 \rangle}= {\frac{3\Lambda}{R^3}}\int_0^R r^{4}dr= {%
\frac{3R^2}{5}}
\end{equation}
and 
\begin{equation}
{\langle G_NM(r)r^{-1} \rangle} = {\frac{3G_NM}{R^6}}\int_0^R r^{4}dr= {%
\frac{3G_NM}{5 R}}
\end{equation}
we recover in the case $\Lambda=0$ the well-known result \cite{kolbturner} 
\begin{equation}
{\frac{3G_NM}{5 R_{\mathrm{max}}}}={\frac{3G_NM}{10 R_{\mathrm{vir}}}},
\end{equation}
i.e., $R_{\mathrm{vir}}=R_{\mathrm{max}}/2$. For a non-vanishing $\Lambda$,
the corresponding equation is 
\begin{equation}
{\frac{3G_NM}{5 R_{\mathrm{max}}}}+{\frac{\Lambda}{10}}R^2_{\mathrm{max}}= {%
\frac{3G_NM}{10 R_{\mathrm{vir}}}}+{\frac{\Lambda}{5}}R^2_{\mathrm{vir}}.
\label{eq:virial}
\end{equation}

Suppose that the mass overdensity contrast compared to the cosmological
average mass density at the maximal (turn-around) radius is $\omega$ ($=5.6$
in the standard case), and assume $\Omega_M+\Omega_\Lambda=1$. Then we can
write $M=4\pi R_{\rm max}^3\omega\rho_M/3$, and this inserted into (\ref
{eq:virial}) gives, by use of (\ref{eq:rhol}) 
\begin{equation}
1+\kappa={\frac{1}{2\mu}}+2\kappa\mu^2  \label{eq:simple}
\end{equation}
where we have introduced $\mu= R_{\mathrm{vir}}/R_{\mathrm{max}}$ ($=0.5$ in
the standard case) and 
\[
\kappa={\frac{1}{\omega}}{\frac{\Omega_\Lambda}{(1-\Omega_\Lambda)(1+z)^3}}. 
\]
This result is written in a somewhat different form than, but agrees with,
Eq.~(26) of \cite{lahav}.
If we assume as a first approximation $\omega=5.6$ as in the $\Lambda=0$
case, and $\Omega_\Lambda=0.7$, we find by solving (\ref{eq:simple})
numerically for $z=0$ that $\mu$ has decreased from 0.5 to around 0.39.

This means that the virialized radius is smaller, which is not unreasonable,
since for a given mass, only more compact clusters can ``survive'' the 
repulsive force from a
positive cosmological constant.

We can improve this analysis somewhat by taking into account the fact that $%
\omega $ also depends on $\Lambda $. It can be seen that this has the effect
of increasing $\omega $ at intermediate redshifts. This increase causes an
increase in $\mu $, meaning that we have overestimated the effect of $%
\Lambda $ above. The effect is, however, small. One can also estimate what
the final density contrast of the cluster will be. To do this we have
obtained an expression comparing the density at maximum expansion with the
density of the universe today. This is given by 
\begin{equation}
\widetilde{\omega }={\frac{9\pi ^{2}}{16}}{\frac{1}{f(a)^{2}}}{\frac{1}{%
(1-\Omega _{\Lambda })}}
\end{equation}
with 
\begin{equation}
f(a)={\frac{3}{2}}\int_{0}^{a}{\frac{da}{\sqrt{(1-\Omega _{\Lambda
})/a+\Omega _{\Lambda }a^{2}}}}.
\end{equation}
The result is a further net relative compression of the clusters 
due to the cosmological constant above the
one given by $\mu $.

Since we have been dealing with an imagined situation where one can compare
between a ``standard'' scenario without a cosmological constant, and one
where $\Lambda \neq 0$, and found only small effects on cluster scales, it
seems difficult, but maybe not excluded, to draw conclusions about the value
of $\Omega _{\Lambda }$ in the real universe where observational
uncertainties have to be taken into account. The effect to search for, is a
tendency for virialized clusters to get smaller and more overdense for a
positive $\Lambda $.

It seems clear, however, that the effects on galactic scales are extremely
tiny and negligible for rotation curves.

\vskip .2cm 
\begin{flushleft}
{\bf\large Acknowledgements}
\vskip .2cm
We wish to thank H. Rubinstein for discussions, and P. Lilje and
M. Rees for informing us about reference \cite{lahav}. This work
was supported in part by the Swedish Natural Science Research Council
(NFR).
\end{flushleft}

\end{document}